\documentstyle[preprint,aps]{revtex}
\begin{document}
%\draft command makes pacs numbers print
\draft
\title{Self-Consistent Velocity Dependent Effective Interactions}
\author{T. Kubo}
\address{Institute for Nuclear Study, University of Tokyo, Tanashi, 
        Tokyo 188, Japan}
\author{H. Sakamoto}
\address{Faculty of Engineering, Gifu University, Gifu 501-11, Japan}
\author{T. Kammuri\cite{byline1}}
\address{Department of Physics, Osaka University, Toyonaka 560, Japan}
\author{T. Kishimoto\cite{byline2}}
\address{Institute of Physics, University of Tsukuba, Ibaraki 305, 
               Japan}
\date{\today}
\maketitle        

\begin{abstract}
The theory of self-consistent effective interactions in nuclei is 
extended for a system with a velocity dependent mean potential. 
By means of the field coupling method, we present a general 
prescription to derive effective interactions which are consistent 
with the mean potential.
For a deformed system with the conventional pairing field, the 
velocity dependent effective interactions are derived as the 
multipole pairing interactions in doubly-stretched coordinates. 
They are applied to the microscopic analysis of the 
giant dipole resonances (GDR's) of ${}^{148,154}Sm$, 
the first excited $2^+$ states of Sn isotopes and the first 
excited $3^-$ states of Mo isotopes. It is clarified that the 
interactions play crucial roles in describing the splitting and 
structure of GDR peaks, in restoring the energy weighted sum 
rule, and in reducing the values of $B(E\lambda)$.
\end{abstract}

% insert suggested PACS numbers in braces on next line
\narrowtext

%%%%%%%%%%%%%%%%%%%%%%%%%%%%%%%%%%%%%%%%%%%%%%%%%%%%%%%
\section{Introduction}
The understanding of nuclear collective excitations has been one of 
the most important subjects in the nuclear many body problem \cite{BM75}. 
In microscopic analyses of such excitations, 
separable multipole interactions have been introduced and applied 
extensively for spherical, deformed and rotating nuclei. 
The separable multipole-multipole interactions originate from long-range 
correlations in particle-hole channels, and the physical meaning of 
these interactions has been clarified by Mottelson 
in terms of the core polarization phenomena \cite{Mo67}. 
On the other hand, the multipole pairing interactions originate from 
short-range correlations in particle-particle channels, 
and there are many works concerning the physical meaning of these 
interactions \cite{BM75,Be67,Ha74,PS77,PGS77,SK90}. 
These interactions have been widely used and have been playing crucial 
roles in the study of nuclear structure, but theoretical foundations 
for the origin of such effective interactions in both the particle-hole 
and particle-particle channels have not been established 
enough from a unified physical picture.
In particular, we consider that it is still an open and 
interesting problem to determine the proper form of the 
deformation dependence of the pairing interaction. 
It is the main purpose of the present paper to report a general 
prescription to derive effective interactions which are consistent 
with the mean potential, and we will present a unified derivation 
of the multipole-multipole interactions in particle-hole channels and 
the multipole pairing interactions in particle-particle channels 
on an equal footing (preliminary reports on this subject can be 
seen in Refs. \cite{KSK88,KSK89}).

Since a nucleus can be regarded as a spatially and energetically 
saturated self-sustained system with a relatively sharp boundary, 
one can assume that the following condition, which will be called 
as {\it nuclear self-consistency} \cite{KMY75,SK89}, is satisfied 
quite accurately: The shape of the mean potential and 
that of the density are the same even when the system undergoes 
collective motions.  The concept of nuclear self-consistency, 
which is much more stringent than the {\it Hartree 
self-consistency}, has played important roles not only in describing 
an equilibrium nuclear shape \cite{Ni55,NTS69} and in deriving 
effective interactions for a spherical system \cite{BM75,Mo67}, 
but also in deriving effective interactions for a deformed system 
\cite{KMY75,SK89}, and higher order effective interactions 
\cite{SK89,KTK83,Ma83}. These interactions have been applied 
successfully to the calculations of the properties of the 
low-lying vibrational states and high-frequency giant 
resonances, etc. 
\cite{KMY75,SK89,KT72,KT76,TWK79,SK88,SK91,MM87,Ai90,MSM90}. 
It is found that the effective interactions derived by the rigorous 
application of the nuclear self-consistency are much more reliable 
than the conventional multipole-multipole interactions 
\cite{BM75,Mo67,KMY75,SK89}.

However, it has been pointed out that the spurious 
velocity dependence in a single particle mean potential, which 
violates the Galilean invariance of the system, should be 
removed \cite{BM75,Be67,PS77,PGS77,SK90}. 
It is well recognized that the 
velocity dependent terms affect the mass parameters of collective 
motions and thereby such quantities as the distribution of 
transition strengths and the absolute value of the energy-weighted 
sum-rule, etc., become unreliable 
\cite{BM75,Be67,PS77,PGS77,SK90,AGG91,CDR91}. 
The realistic nuclear potential in fact contains velocity dependent 
terms such as the BCS pairing field, $\vec l \cdot \vec s$ and 
$\vec l^2$ terms in the Nilsson model, etc. Some parts of them are 
responsible for violating the Galilean invariance and the classical 
sum-rule, and therefore these symmetries have to be restored 
properly. From this point of view, Pyatov {\it et al.} developed 
a simple and powerful method to introduce additional interactions 
which restore such broken symmetries in RPA order 
\cite{PS77,PGS77}. Since then, symmetry restoring effective 
interactions have been studied by several authors 
\cite{PS77,PGS77,SK90,CKC84,NF88,CL88,CFL92,NFD95}. 
In this paper, by investigating the coupling between 
collective and single-particle degrees of freedom, 
and by using the concept of {\it nuclear self-consistency} and 
{\it local Galilean invariance} of the system as important guiding 
principles, we will present a systematic method to derive 
self-consistent effective interactions in nuclei.

In sect.\ref{sect2}, the field coupling method \cite{BM75,SK89} 
is applied to a spherical system with a velocity dependent 
potential. In this framework, we derive self-consistent 
velocity dependent effective interactions by estimating 
the coupling between the collective displacement of nucleons 
and the mean field. 
In sect.\ref{sect3}, we extend the field coupling method to a 
deformed system and present a simple derivation of 
the doubly-stretched multipole-multipole and multipole-pairing 
interactions.
In sect.\ref{sect4}, we investigate some fundamental properties 
of these self-consistent effective interactions. 
It is shown to be essential to express effective interactions in 
doubly-stretched coordinates for restoring 
some broken symmetries and also for the natural description 
of GDR's in deformed nuclei. 
In sect.\ref{sect5}, we report the results of numerical 
calculations in RPA of the GDR's of ${}^{148,154}Sm$, the first 
excited $2^+$ states of Sn isotopes and the first excited $3^-$ 
states of Mo isotopes, by using the self-consistent velocity 
dependent effective interactions.

%%%%%%%%%%%%%%%%%%%%%%%%%%%%%%%%%%%%%%%%%%%%%%%%%%%%%%%
\section{General Framework of Field Coupling Method}\label{sect2}
%---------------------------------------------------------------------
\subsection{Field coupling between collective distortion and mean 
            potential}\label{Field}
The method to study collective motion that arises from the action of 
the field coupling in a system with a degenerate one-particle 
excitation has been well developed by Bohr and Mottelson \cite{BM75}. 
It provides a self-consistent method to construct the relevant 
effective interaction by identifying the field coupling as the 
Hartree field of the interaction.
Based on the method, which will be referred to 
as the {\it field coupling method}, the theory of self-consistent 
effective interactions in nuclei has been developed 
\cite{KMY75,SK89,KTK83}. As a result, it has been shown for example 
that the conventional multipole interaction model must be improved to 
satisfy the nuclear self-consistency in deformed nuclei, resulting in 
the doubly-stretched multipole interaction model.
Here we will extensively apply the method to a system with the 
velocity dependent mean potential. 
Some examples of the velocity dependent field coupling were 
discussed in Ref.\cite{BM75} in connection with the analysis of 
the high-frequency quadrupole modes and of the center-of-mass mode. 

In a nuclear system, particle motions and collective motions are 
essentially coupled with each other in order to achieve 
self-consistency between the potential and the density distribution. 
If a collective mode excites in a nucleus, the 
positions and momenta of nucleons in the system are displaced 
by the mode accordingly. 
Then the corresponding change in the nuclear density distribution 
gives rise to a violation of the self-consistency settled before 
the displacements are switched on. 
In this way additional field couplings are induced in the system 
in order to restore the nuclear self-consistency.

Let us start by considering a $2^\lambda$-pole collective shape 
oscillation mode of a spherical nucleus.  We assume that this mode 
is characterized by the collective displacement of the nucleonic 
field variable as
\begin{equation}\label{dr}
\vec r\ \to \ \vec r+\delta \vec r, \ \ \ \ \ 
\delta \vec r=-\sum\limits_\mu \alpha_{\lambda \mu}^*
\vec \nabla Q_{\lambda \mu}\ \ \ \ \ (\lambda \ge 1),
\end{equation}
where $\alpha_{\lambda \mu}$ are the collective amplitudes.
The collective velocity field associated with this mode is 
expressed as
\begin{equation}\label{v}
\vec v(\vec r)=-\sum\limits_\mu \dot 
\alpha_{\lambda \mu}^*\vec \nabla Q_{\lambda \mu},
\end{equation}
which is irrotational and incompressible.  We will require that 
nucleonic velocities entering into the velocity-dependent 
single-particle potential are to be measured relative to the 
collective velocity $\vec v(\vec r)$ \cite{BM75} so that the 
potential is to be invariant under the local Galilean transformation 
\begin{equation}
\vec p\ \to \ \vec p+\delta \vec p, \ \ \ \ \ 
\delta \vec p=M\vec v(\vec r).
\end{equation}
Then the variation of the average one-body potential $\delta V$ 
and that of the density distribution $\delta \rho$ produced by the 
oscillation are determined from the conditions
\begin{eqnarray}\label{disp}
V(\vec r+\delta \vec r,\vec p+\delta \vec p)
 &=& V_0(\vec r,\vec p), \nonumber \\
\rho(\vec r+\delta \vec r,\vec p+\delta \vec p)
 &=& \rho_0(\vec r,\vec p),
\end{eqnarray}
where $V_0(\vec r,\vec p)$ and $\rho_0(\vec r,\vec p)$ are the 
potential and the density at the original equilibrium point in the 
phase space and are assumed to be spherical, while 
$V(\vec r,\vec p)$ and $\rho(\vec r,\vec p)$ include the effect 
of the oscillation. In this paper we assume that these quantities 
expressed in the phase space are well defined by using some 
appropriate semiclassical method such as 
the Wigner transformation \cite{Wi32,RS80,BS80,DK87}.

The conditions of Eq.(\ref{disp}) provide relations between 
$\delta V$ and $\delta \rho$ through the displacement vectors 
$\delta \vec r$ and $\delta \vec p$ as 
\begin{eqnarray}
V(\vec r,\vec p)&=&V_0(\vec r,\vec p)+\delta V(\vec r,\vec p), 
\nonumber \\
\rho(\vec r,\vec p)&=&\rho_0(\vec r,\vec p)+\delta \rho(\vec r,\vec p),
\end{eqnarray}
with
\begin{eqnarray}\label{delV}
\delta V(\vec r,\vec p)&=&\delta V_r+\delta V_p=-\delta \vec r\cdot 
\vec \nabla V_0-\delta \vec p\cdot \vec \nabla _pV_0, \nonumber \\
\delta \rho(\vec r,\vec p)&=&\delta \rho_r+\delta \rho_p=-\delta 
\vec r\cdot \vec \nabla \rho_0-\delta \vec p\cdot \vec \nabla _p\rho_0,
\end{eqnarray}
where notations are defined by 
\begin{equation}
\vec \nabla \equiv \left({{\partial \over {\partial x}},
{\partial \over {\partial y}},{\partial \over {\partial z}}} \right),
\ \ \vec \nabla _p\equiv \left({{\partial \over {\partial p_x}},
{\partial \over {\partial p_y}},{\partial \over {\partial p_z}}} 
\right).
\end{equation} 
In Eq.(\ref{delV}), for simplicity, only the leading order terms in 
$\alpha_{\lambda \mu}$ and $\dot \alpha _{\lambda \mu}$ 
are retained assuming that the vibrational amplitudes are small. 
Inclusion of the non-linear field coupling coming from 
higher-order terms in $\alpha_{\lambda \mu}$ has been performed in 
Ref.\cite{SK89} resulting in higher-order (many-body) effective 
interactions. The $\delta V_r$ and $\delta V_p$ can be expressed 
in the standard form of the field coupling \cite{BM75} as 
\begin{equation}\label{F}
\delta V_r=\kappa_\lambda \sum\limits_\mu 
\alpha _{\lambda \mu}^*F_{\lambda \mu}, \ \ \ \ \ 
F_{\lambda \mu}={1 \over {\kappa_\lambda}}\vec \nabla Q_{\lambda \mu}
\cdot \vec \nabla V_0, 
\end{equation}
\begin{equation}\label{Ftild}
\delta V_p=\tilde \kappa_\lambda \sum\limits_\mu 
\dot \alpha _{\lambda \mu }^*\tilde F_{\lambda \mu}, \ \ \ \ \ 
\tilde F_{\lambda \mu}={M \over {\tilde \kappa_\lambda}}\vec 
\nabla Q_{\lambda \mu}\cdot \vec \nabla_pV_0.
\end{equation}

Now we consider that the field couplings $\delta V_r$ and 
$\delta V_p$ are produced as the Hartree field of two-body 
interactions 
\begin{equation}
H_{int}={\kappa_\lambda \over 2}\sum\limits_\mu 
F_{\lambda \mu}^{\dag} F_{\lambda \mu}
+{\tilde \kappa_\lambda \over 2}\sum\limits_\mu 
\tilde F_{\lambda \mu}^{\dag} \tilde F_{\lambda \mu}
\equiv H_\lambda^{(r)}+H_\lambda^{(p)},
\end{equation}
and introduce self-consistency conditions as 
\begin{equation}\label{self}
\alpha_{\lambda \mu}^*=A<F_{\lambda \mu}^{\dag}>, \ \ \ \ \ 
\dot \alpha_{\lambda \mu}^*=A<\tilde F_{\lambda \mu}^{\dag}>.
\end{equation}
Here the average of a one-body operator with respect to the modified 
ground state $|>$ corresponding to the density $\rho$ is calculated 
as 
\begin{equation}
A<F>=<\sum\limits_{i=1}^A F(i)>=\int F\rho (\vec r,\vec p)d^3rd^3p,
\end{equation}
while that with respect to the original ground state $|0>$ 
corresponding to the density $\rho_0$ is given by
\begin{equation}
A<F>_0=<\sum\limits_{i=1}^A F(i)>_0
=\int F\rho_0 (\vec r,\vec p)d^3rd^3p.
\end{equation}
Equation (\ref{self}) can be calculated as
\begin{eqnarray}
\alpha_{\lambda \mu}^*
&=&\int F_{\lambda \mu}^{\dag} \delta \rho_r \ d^3rd^3p \nonumber\\
&=&\alpha_{\lambda \mu}^*\int F_{\lambda \mu}^{\dag}
   \vec \nabla Q_{\lambda \mu}\cdot \vec \nabla \rho_0 \ d^3rd^3p 
   \nonumber\\
&=&-\alpha_{\lambda \mu}^*A<(\vec \nabla F_{\lambda \mu}^{\dag}\cdot 
   \vec \nabla Q_{\lambda \mu })>_0, \\
\dot \alpha_{\lambda \mu}^*
&=&\int \tilde F_{\lambda \mu }^{\dag} 
   \delta \rho_p \ d^3rd^3p \nonumber\\
&=&M \dot \alpha_{\lambda \mu}^*
   \int \tilde F_{\lambda \mu}^{\dag}\vec \nabla Q_{\lambda \mu}\cdot 
   \vec \nabla_p\rho _0 \ d^3rd^3p  \nonumber\\
&=&-M \dot \alpha _{\lambda \mu}^*A<(\vec \nabla_p
   \tilde F_{\lambda \mu}^{\dag}\cdot \vec \nabla Q_{\lambda \mu})>_0,
\end{eqnarray}
where we have assumed the time reversal invariance of the density 
$\rho_0$ which guarantees 
\begin{equation}
\int F_{\lambda \mu}^{\dag}\delta \rho _p \ d^3rd^3p
\propto A<\vec \nabla _pF_{\lambda \mu}^{\dag}
\cdot \vec \nabla Q_{\lambda \mu}>_0=0,
\end{equation}
\begin{equation}
\int \tilde F_{\lambda \mu }^{\dag}\delta \rho _r \ d^3rd^3p
\propto A<\vec \nabla \tilde F_{\lambda \mu}^{\dag}
\cdot \vec \nabla Q_{\lambda \mu}>_0=0.
\end{equation}
We thus obtain
\begin{equation}\label{k}
\kappa_\lambda =-A<\vec \nabla \left( 
\vec \nabla Q_{\lambda \mu}^{\dag}\cdot \vec \nabla V_0 \right)
\cdot \vec \nabla Q_{\lambda \mu}>_0,
\end{equation}
\begin{equation}\label{ktild}
\tilde \kappa_\lambda =-M^2A<\vec \nabla _p\left( 
\vec \nabla Q_{\lambda \mu}^{\dag}\cdot \vec \nabla _pV_0 
\right)\cdot \vec \nabla Q_{\lambda \mu}>_0.
\end{equation}
Since $Q_{\lambda \mu}$ does not depend on the momentum $\vec p$, 
$\tilde F_{\lambda \mu}$ of Eq.(\ref{Ftild}) and 
$\tilde \kappa_\lambda$ of Eq.(\ref{ktild}) can be expressed in terms 
of the Poisson bracket as 
\begin{equation}\label{Ftild2}
\tilde F_{\lambda \mu}={M \over {\tilde \kappa_\lambda}}
\{ Q_{\lambda \mu},V_0 \},
\end{equation}
\begin{equation}\label{ktild2}
\tilde \kappa_\lambda =-M^2<\{ Q_{\lambda \mu},
\{ Q_{\lambda \mu}^{\dag},V_0 \}\}>_0.
\end{equation}

In the above treatment, we have assumed the potential $V_0$ can be 
expressed as a sum over individual particles: 
$V_0=\sum\limits_{i=1}^A V_0(\vec r_i,\vec p_i)$. 
However the monopole pairing potential
\begin{equation}\label{Vpair}
V_{pair}=-{\Delta \over 2}(P_0^{\dag}+P_0),\ \ \ \ \ 
P_0^{\dag}\equiv \sum\limits_\gamma c_\gamma^{\dag} 
c_{\tilde \gamma}^{\dag}
=2\sum\limits_{\gamma > 0}c_\gamma^{\dag} 
c_{\tilde \gamma}^{\dag}
\end{equation}
is not expressed in this form.  Here $c_\gamma^{\dag}$ creates a 
nucleon in the state $\gamma$, $\tilde \gamma$ is a time-reversed 
state of $\gamma$, and $\Delta$ is a gap parameter. For such 
a potential we will replace the Poisson brackets in 
Eqs.(\ref{Ftild2}) and (\ref{ktild2}) by the corresponding 
commutation relations as
\begin{equation}\label{Ftild3}
\tilde F_{\lambda \mu}={M \over {i\hbar\tilde \kappa_\lambda}} 
[Q_{\lambda \mu},V_0],
\end{equation}
\begin{equation}\label{ktild3}
\tilde \kappa_\lambda =-\left({M \over \hbar}\right)^2
<[Q_{\lambda \mu},[V_0,Q_{\lambda \mu}^{\dag}]\ ]>_0.
\end{equation}
Then the corresponding part of the effective interaction becomes
\begin{equation}\label{Hsym}
H_\lambda^{(p)}=-{1 \over 2}\sum\limits_\mu 
{1 \over {<[Q_{\lambda \mu},[V_0,Q_{\lambda \mu}^{\dag}]\ ]>_0}}
[Q_{\lambda \mu},V_0]^{\dag}[Q_{\lambda \mu},V_0].
\end{equation}
This type of separable effective interactions have been discussed 
by several authors in connection with the symmetry-restoring 
treatment of the nuclear Hamiltonian 
\cite{PS77,PGS77,SK90,CKC84,NF88,CL88,CFL92,NFD95}.
%---------------------------------------------------------------------
\subsection{Two simple examples of self-consistent effective 
                  interactions}
It is worthwhile to briefly review the derivation of the 
self-consistent effective interactions in the case of a simple mean 
potential such as a harmonic oscillator potential \cite{SK89} 
or a pairing potential \cite{SK90}.
%+++++++++++++++++++++++++++++++++++++++++++++++++++++++++++++++++++++
\subsubsection{Spherical harmonic oscillator potential}
Let us consider a simple situation that the equilibrium potential 
$V_0$ is given by the spherical harmonic oscillator potential 
\begin{equation}
V_0=V_{ho}(r)={1 \over 2}M\omega _0^2r^2.
\end{equation}
In this case, we do not have to consider the velocity dependent field 
coupling since the potential does not have any velocity 
dependence. From Eqs.(\ref{F}) and (\ref{k}) we obtain 
\begin{equation}
F_{\lambda \mu}={1 \over \kappa_\lambda} M\omega_0^2 \ 
\vec r\cdot \vec \nabla Q_{\lambda \mu}
={1 \over \kappa_\lambda} \lambda M\omega_0^2 \ Q_{\lambda \mu}, 
\end{equation}
\begin{eqnarray}
\kappa_\lambda
&=&-\lambda M\omega_0^2 A<\vec \nabla Q_{\lambda \mu}^{\dag} 
   \cdot \vec \nabla Q_{\lambda \mu}>_0  \nonumber\\
&=&-M\omega_0^2 {{\lambda^2 (2\lambda +1)} \over {4\pi}} 
   A<r^{2\lambda -2}>_0, 
\end{eqnarray}
where we used the relations
\begin{eqnarray}
& \vec r\cdot \vec \nabla Q_{\lambda \mu }=\lambda Q_{\lambda \mu },
  \ \ \ \Delta Q_{\lambda \mu }=0, & \nonumber\\
& \Delta (Q_{\lambda \mu}^{\dag} Q_{\lambda \mu})
  ={{2\lambda (2\lambda +1)} \over {4\pi}} r^{2\lambda -2}. &
\end{eqnarray}
Thus the self-consistent effective interaction coincides with the 
conventional multipole-multipole interaction 
\begin{equation}
H_\lambda^{(r)}=-{1 \over 2}\chi_\lambda^{self}\sum\limits_\mu 
Q_{\lambda \mu}^{\dag} Q_{\lambda \mu}
\end{equation}
with the self-consistent strength
\begin{equation}
\chi_\lambda^{self}={{4\pi} \over {2\lambda +1}}
{{M\omega _0^2} \over {A<r^{2\lambda -2}>_0}}.
\end{equation}
%+++++++++++++++++++++++++++++++++++++++++++++++++++++++++++++++++++++
\subsubsection{Pairing potential}
Let us introduce a one-body operator 
\begin{equation}
O=\sum\limits_{\alpha \beta}<\alpha |O|\beta >c_\alpha^{\dag} c_\beta ,
\end{equation}
whose time-reversal property is assumed to be 
\begin{equation}
\hat TO\hat T^{-1}=(-)^TO^{\dag}
\end{equation}
or equivalently 
\begin{equation}
<\tilde \alpha |O|\tilde \beta >=<\hat T\alpha |O|\hat T\beta >
=(-)^T<\beta |O|\alpha >,
\end{equation}
where $\hat T$  denotes the time-reversal operator, and $(-)^T$ 
is a short-hand notation for the time-reversal phase and is either 
$+1$ or $-1$. The commutation relation between the operator O and 
the pairing field $V_{pair}$ is given by
\begin{equation}\label{comOV}
[O,V_{pair}]={{1+(-)^T} \over 2}\Delta\sum\limits_{\alpha \beta} 
<\alpha |O|\tilde \beta >(c_\alpha^{\dag} c_\beta^{\dag} 
-c_{\tilde \beta }c_{\tilde \alpha }).
\end{equation}
 From this relation, it is easily recognized that the pairing field 
satisfies the translational invariance but violates the Galilean 
invariance:
\begin{equation}
[\vec P,\ V_{pair}]=0,
\end{equation}
\begin{equation}
[\vec R,V_{pair}]=\Delta\sum\limits_{\alpha \beta} 
<\alpha |\vec R|\tilde \beta >(c_\alpha^{\dag} c_\beta^{\dag} 
-c_{\tilde \beta}c_{\tilde \alpha})
\end{equation}
with
\begin{equation}
{\vec P}= \sum_{j=1}^A {\vec p_j}= \sum_{j=1}^A 
\left(-i \hbar \vec \nabla_j \right), 
\ \ \ \ \ 
{\vec R}= {1 \over A} \sum_{j=1}^A {\vec r_j}.
\end{equation}
More generally, if O is an operator depending on coordinate 
variables, then Eq.(\ref{comOV}) means that the monopole pairing 
field is not invariant under a local Galilean transformation.
In this sense the pairing field is considered to be a velocity 
dependent mean potential \cite{BM75}. For a system with such a 
velocity dependent potential, there arises a velocity dependent 
field coupling as is discussed in subsect.\ref{Field}. 

Let us investigate the interaction of Eq.(\ref{Hsym}) in more detail. 
By substituting $V_{pair}$ into $V_0$ and by identifying the 
original ground state $|0>$ with the BCS vacuum 
state, the basic quantities in Eqs.(\ref{Ftild3}) and (\ref{ktild3}) 
are expressed as
\begin{equation}
[Q_{\lambda \mu},V_{pair}]=-\Delta(P_{\lambda \mu}^{\dag} 
-P_{\widetilde{\lambda \mu}}),
\end{equation}
\begin{eqnarray}
<[Q_{\lambda \mu}
&,&[V_{pair},Q_{\lambda \mu}^{\dag}]\ ]>_0 \nonumber\\
&=&\Delta^2\sum\limits_{\alpha \beta}({1 \over {E_\alpha}}
   +{1 \over {E_\beta}})|<\alpha |Q_{\lambda \mu}|\beta >|^2,
\end{eqnarray}
where 
\begin{equation}
P_{\lambda \mu}^{\dag} \equiv \sum\limits_{\alpha \beta}
<\alpha |Q_{\lambda \mu}|\beta >c_\alpha^{\dag} 
c_{\tilde \beta}^{\dag},
\end{equation}
and we have used the familiar relation between the coefficients 
of Bogoliubov-Valatin transformation and the quasiparticle energy: 
$u_\alpha v_\alpha =\Delta/2E_\alpha$.
Thus the velocity dependent effective interaction of 
Eq.(\ref{Hsym}) can be obtained as
\begin{equation}
H_\lambda^{(p)}=-{1 \over 8}G_\lambda^{self}\sum\limits_\mu 
(P_{\lambda \mu}^{\dag}-P_{\widetilde{\lambda \mu}})
(P_{\lambda \mu}-P_{\widetilde{\lambda \mu}}^{\dag})
\end{equation}
with
\begin{equation}\label{Gself}
G_\lambda^{self}\equiv 1 \left/  \sum\limits_{\alpha \beta } 
{1 \over 4}({1 \over {E_\alpha}}+{1 \over {E_\beta}})
|<\alpha |Q_{\lambda \mu}|\beta >|^2 \right. .
\end{equation}
This interaction is a natural extension of the dipole-pairing 
interaction obtained by Bohr and Mottelson \cite{BM75} and 
Pyatov and Salamov \cite{PS77} to a general $2^\lambda$-pole mode. 
Equation (\ref{Gself}) has the same structure as that of the Belyaev 
identity \cite{Be67,Ha74}. Further discussions for this interaction 
can be found in Refs.\cite{SK90,SK91}.

%%%%%%%%%%%%%%%%%%%%%%%%%%%%%%%%%%%%%%%%%%%%%%%%%%%%%%%
\section{Self-Consistent Effective Interactions in Deformed Nuclei}
\label{sect3}

Now, let us apply the procedure of sect.\ref{sect2} to a deformed 
system. For simplicity, we assume that the main part of the 
equilibrium potential is described as a deformed harmonic oscillator 
with frequencies $\omega_x$, $\omega_y$ and $\omega_z$. In terms of 
the doubly-stretched coordinates \cite{KMY75,SK89} defined by 
\begin{equation}
x''={{\omega_x} \over {\omega_0}}x,\ \ 
y''={{\omega_y} \over {\omega_0}}y,\ \ 
z''={{\omega _z} \over {\omega _0}}z,
\end{equation}
the deformed harmonic oscillator potential can be expressed in a 
spherical form as 
\begin{eqnarray}
V_{ho}(\vec r)&=&{1 \over 2}M\left( {\omega_x^2x^2+\omega_y^2y^2
                 +\omega_z^2z^2} \right) \nonumber\\
              &=&{1 \over 2}M\omega_0^2(r'')^2=V_{ho}(r'').
\end{eqnarray}

As one of the possible and plausible collective shape oscillation 
mode in the deformed system, we will introduce a doubly-stretched 
$2^\lambda$-pole mode characterized by 
\begin{equation}\label{dr2}
\vec r\:''\ \to \ \vec r\:''+\delta \vec r\:'', \ \ \ 
\delta \vec r\:''=-\sum\limits_\mu \alpha_{\lambda \mu}^*
\vec \nabla'' Q''_{\lambda \mu}\ \ \ (\lambda \ge 1),
\end{equation}
with 
\begin{equation}
\vec \nabla'' \equiv \left({{\partial \over {\partial x''}},
{\partial \over {\partial y''}},{\partial \over {\partial z''}}} 
\right),\ \ \ Q''_{\lambda \mu}\equiv Q_{\lambda \mu}(\vec r\:''),
\end{equation} 
rather than the conventional $2^\lambda$-pole mode characterized by 
Eq.(\ref{dr}). In fact, as discussed in Ref.\cite{SK89} based on the 
Thomas-Fermi theory, the doubly-stretched mode represents great 
improvements over the conventional mode in the sense that it 
satisfies (i) the constancy of the Fermi energy (which is equivalent 
to the saturation condition), (ii) the separation of the 
center-of-mass motion, (iii) the condition for a fluctuation around 
the deformed equilibrium shape, and (iv) the self-consistency between 
the nucleonic density and the potential, etc. The collective velocity 
field associated with this mode is expressed as
\begin{equation}
\vec v\:''(\vec r\:'')=-\sum\limits_\mu \dot 
\alpha_{\lambda \mu}^*\vec \nabla'' Q''_{\lambda \mu},
\end{equation}
which is a natural extension of the irrotational and incompressible 
flow of Eq.(\ref{v}).  

Now we will extensively apply the requirement of the local Galilean 
invariance to the deformed system. Namely, we will require that 
in terms of the doubly-stretched coordinates nucleonic velocities 
entering into the velocity-dependent single-particle potential are 
to be measured relative to the collective velocity 
$\vec v\:''(\vec r\:'')$ so that the 
potential is to be invariant under the transformation 
\begin{equation}\label{dp2}
\vec p\:''\ \to \ \vec p\:''+\delta \vec p\:'', \ \ \ \ \ 
\delta \vec p\:''=M\vec v\:''(\vec r\:''),
\end{equation}
where the doubly-stretched momentums are introduced as
\begin{equation}
p''_x={{\omega_0} \over {\omega_x}}p_x,\ \ \ 
p''_y={{\omega_0} \over {\omega_y}}p_y,\ \ \ 
p''_z={{\omega_0} \over {\omega_z}}p_z.
\end{equation}
Then by use of the relations $d^3r=d^3r''$ and $d^3p=d^3p''$, which 
are equivalent to the volume conservation condition 
$\omega_x\omega_y\omega_z=\omega_0^3$, we can follow the similar 
procedure as given in sect.\ref{sect2} to derive self-consistent 
effective interactions in deformed nuclei. To do this, 
in the present case, we should understand that all the $\vec r$, 
$\vec p$, $\vec \nabla$ and $\vec \nabla_p$ appearing in 
sect.\ref{sect2} are to be replaced by the corresponding 
$\vec r\:''$, $\vec p\:''$, $\vec \nabla''$ and $\vec \nabla''_p$ 
with 
\begin{equation}
\vec \nabla''_p \equiv \left( {{\partial \over {\partial p_x''}},
{\partial \over {\partial p_y''}},
{\partial \over {\partial p_z''}}} \right).
\end{equation}
For example, the field coupling induced by the collective 
displacements of Eqs.(\ref{dr2}) and (\ref{dp2}) becomes 
\begin{equation}
V(\vec r\:'',\vec p\:'')=V_0(\vec r\:'',\vec p\:'')
+\delta V(\vec r\:'',\vec p\:'') \equiv V''_0+\delta V''
\end{equation}
with
\begin{eqnarray}
\delta V''
&=&\delta V''_r+\delta V''_p \nonumber\\
&=&\kappa_\lambda \sum\limits_\mu 
   \alpha _{\lambda \mu}^*F''_{\lambda \mu},
   +\tilde \kappa_\lambda \sum\limits_\mu 
   \dot \alpha _{\lambda \mu}^*\tilde F''_{\lambda \mu}, \\
F''_{\lambda \mu}&=&{1 \over {\kappa_\lambda}}\vec \nabla'' 
Q''_{\lambda \mu}\cdot \vec \nabla\:'' V''_0,  \\
\tilde F''_{\lambda \mu}&=&{M \over {\tilde \kappa_\lambda}}
\vec \nabla'' Q''_{\lambda \mu}\cdot \vec \nabla\:''_pV''_0
={M \over {\tilde \kappa_\lambda}}\{ Q''_{\lambda \mu},V''_0 \}.
\end{eqnarray}
 From the self-consistency conditions
\begin{equation}
\alpha_{\lambda \mu}=A<F''_{\lambda \mu}>, \ \ \ \ \ 
\dot \alpha_{\lambda \mu}=A<\tilde F''_{\lambda \mu}>,
\end{equation}
the coupling strengths are determined as 
\begin{eqnarray}
\kappa_\lambda &=& -A<\vec \nabla'' \left( 
\vec \nabla'' Q''_{\lambda \mu}{}^{\dag}\cdot 
\vec \nabla'' V''_0 \right)
\cdot \vec \nabla'' Q''_{\lambda \mu}>_0,     \\
\tilde \kappa_\lambda &=& -M^2A<\vec \nabla''_p\left( 
\vec \nabla'' Q''_{\lambda \mu}{}^{\dag}\cdot \vec \nabla''_pV''_0 
\right)\cdot \vec \nabla'' Q''_{\lambda \mu}>_0  \nonumber\\ 
&=&-M^2<\{ Q''_{\lambda \mu},
\{ Q''_{\lambda \mu}{}^{\dag},V''_0 \}\}>_0,
\end{eqnarray}
and the self-consistent effective interaction to be used in the 
deformed nucleus is given as
\begin{eqnarray}\label{Hint}
H_{int}&=&{\kappa_\lambda \over 2}\sum\limits_\mu 
F''_{\lambda \mu}{}^{\dag} F''_{\lambda \mu}
+{\tilde \kappa_\lambda \over 2}\sum\limits_\mu 
\tilde F''_{\lambda \mu}{}^{\dag} \tilde F''_{\lambda \mu} \nonumber\\
&\equiv& H_\lambda^{(r)}+H_\lambda^{(p)}.
\end{eqnarray}
For the case of $V''_0=V_{ho}(r'')$, the effective 
interaction becomes 
\begin{equation}
H_\lambda^{(r)}=-{1 \over 2}\chi_\lambda^{self}\sum\limits_\mu 
Q''_{\lambda \mu}{}^{\dag} Q''_{\lambda \mu}
\end{equation}
with
\begin{equation}
\chi_\lambda^{self}={{4\pi} \over {2\lambda +1}}
{{M\omega _0^2} \over {A<(r'')^{2\lambda -2}>_0}},
\end{equation}
and in the presence of the pairing field an additional effective 
interaction is derived as 
\begin{equation}\label{Hpair}
H_\lambda^{(p)}=-{1 \over 8}G_\lambda^{self}\sum\limits_\mu 
(P''_{\lambda \mu}{}^{\dag}-P''_{\widetilde{\lambda \mu}})
(P''_{\lambda \mu}-P''_{\widetilde{\lambda \mu}}{}^{\dag})
\end{equation}
with
\begin{equation}
G_\lambda^{self}\equiv 1 \left/  \sum\limits_{\alpha \beta } 
{1 \over 4}({1 \over {E_\alpha}}+{1 \over {E_\beta}})
|<\alpha |Q''_{\lambda \mu}|\beta >|^2 \right. .
\end{equation}
These interactions play crucial roles in restoring broken 
symmetries of the system in RPA order, the detail of which will be 
discussed in the next section.

%%%%%%%%%%%%%%%%%%%%%%%%%%%%%%%%%%%%%%%%%%%%%%%%%%%%%%%
\section{Fundamental Properties of Self-Consistent Effective 
         Interactions}\label{sect4}
%---------------------------------------------------------------------
\subsection{Translational invariance of a deformed system}
The consistency of the residual interaction with the shell model 
potential and the method to restore the translational invariance 
of a nuclear many-body system was discussed by Pyatov and 
Salamov \cite{PS77} for the case of the spherical oscillator 
potential. Here we will briefly examine this problem in a 
deformed system.

Let us consider the oscillator Hamiltonian for the deformed nucleus 
\begin{equation}\label{defHO}
H_0 = \sum_{i=1}^A \left\{ {p_i^2 \over 2M}  
    + V_{ho}(r''_i) \right\}.
\end{equation}         
Since $H_0$ is a local one-body Hamiltonian, it breaks the 
translational invariance of the system.  In fact, we have
\begin{equation}
[ H_0 , {\vec P}''] = i \hbar AM \omega_0^2 {\vec R}''
\end{equation}
with
\begin{equation}
{\vec P}'' = \sum_{j=1}^A {\vec p_j}{}'' = \sum_{j=1}^A 
\left(-i \hbar \vec \nabla_j'' \right), 
\ \ \ \ \ 
{\vec R}'' = {1 \over A} \sum_{j=1}^A {\vec r_j}{}''.
\end{equation}
To recover the translational invariance, we need a counter term
which cancels the above commutation relation. For a spherical
nucleus, as is well known, such a counter term comes from the 
conventional dipole interaction. Now we will show that for the 
deformed nucleus, the doubly-stretched dipole interaction plays 
the role to recover the translational invariance of the system.

First of all, for the doubly-stretched dipole interaction
\begin{equation}
V_{\lambda=1}= -{\chi_1 \over 2} \sum_{ij} \left( Q_1''(i)
     \cdot Q_1''(j) \right),
\end{equation}
we can verify
\begin{equation}
[ V_{\lambda=1}, {\vec P}'' ] = -i \hbar AM \omega_0^2
{\chi_1 \over \chi_1^{self}}{\vec R}''.
\end{equation}
Then the total Hamiltonian $H = H_0 + V_{\lambda=1}$ satisfies 
\begin{equation}
[ H, {\vec P}''] = i \hbar AM \omega_0^2
   \left( 1 - {\chi_1 \over \chi_1^{self}} \right) {\vec R}''.
\end{equation}
Therefore if the strength $\chi_1$ of the doubly-stretched dipole 
interaction is set equal to its self-consistent value 
$\chi_1^{self}=4\pi M\omega_0^2/3A$, we obtain $[H, {\vec P}''] = 0$ 
exactly. In this case, the total Hamiltonian can be expressed as
\begin{equation}
H= \sum_{i=1}^A {p_i^2 \over 2M} + { M\omega_0^2 \over 4A} 
\sum_{i,j=1}^A \left| {\vec r_i}{}''-{\vec r_j}{}'' \right|^2 ,
\end{equation}
which explicitly guarantees the translational invariance of the 
deformed system.

%---------------------------------------------------------------------
\subsection{Restoration of the local Galilean invariance}

As is discussed in the previous sections, generally a 
phenomenological potential, comprising velocity dependent terms 
such as the $\vec l{}^2$ term, the $\vec l \cdot \vec s$ terms and 
the pairing field, etc., does not commute with arbitrary coordinate 
operators. If we chose the operator to be the multipole operator, 
such a situation is expressed from Eq.(\ref{Ftild3}) as 
\begin{equation}
[Q_{\lambda \mu}'' \ , V_0(\vec r, \vec p)]
={{i\hbar\tilde \kappa_\lambda} \over M}\tilde F''_{\lambda \mu}
\ne 0, 
\end{equation}
which means that the potential $V_0$ violates the local Galilean 
invariance under the collective multipole oscillation. 
Here the doubly-stretched operators are used for deformed nuclei. 
For spherical nuclei, of course we can omit the double primes.  

Now along the line of the general method of restoring the broken 
symmetry \cite{PS77,CKC84}, we will show that the 
self-consistent velocity dependent effective interaction, i.e., 
$H_\lambda^{(p)}$ of Eq.(\ref{Hint}), plays the role to restore 
the local Galilean invariance of the system under the random phase 
approximation (RPA). In the RPA order we can verify 
\begin{equation}
[Q_{\lambda \mu}'' \ , H_\lambda^{(p)} ]_{RPA}
={\tilde \kappa_\lambda \over 2} \left\{
[Q_{\lambda \mu}'' \ , \tilde F''_{\lambda \mu}{}^{\dag}]_{RPA} , 
\tilde F''_{\lambda \mu} \right\}_+,
\end{equation}
with
\begin{equation}
[Q_{\lambda \mu}'' \ , \tilde F''_{\lambda \mu}{}^{\dag}]_{RPA}
={M \over {i\hbar\tilde \kappa_\lambda}}
<[Q''_{\lambda \mu} \ , [Q''_{\lambda \mu}{}^{\dag} \ , V_0]]>_0,
=- {i\hbar \over M}.
\end{equation}
Thus we obtain 
\begin{equation}
[Q_{\lambda \mu}'' \ , V_0+H_\lambda^{(p)} ]_{RPA}=0.
\end{equation}

%---------------------------------------------------------------------
\subsection{Simple model analysis of GDR of normal nuclei}

It is worthwhile to point out that the doubly-stretched 
interaction model is powerful and plausible also for the 
description of some iso-vector modes. 
In confirmation of it, let us briefly review the simple model 
analysis of the splitting of GDR in an axially symmetric deformed 
nucleus \cite{SK89}. 
The model Hamiltonian is assumed to be 
\begin{equation}
H=H_0+V_{\lambda=1}^{(T=1)},
\end{equation}
where $H_0$ is the deformed oscillator Hamiltonian of 
Eq.(\ref{defHO}) while $V_{\lambda=1}^{(T=1)}$ is a residual 
iso-vector dipole interaction.
Here we will parametrize the shape of the nuclear potential as
\begin{equation}
\omega_\bot=\omega_x=\omega_y=\omega_0(\varepsilon)
(1+\varepsilon /3),\ \ \ 
\omega_z=\omega_0(\varepsilon)(1-2\varepsilon /3)
\end{equation}
with
\begin{equation}
\omega_0(\varepsilon) = \stackrel{\circ}{\omega}_0
(1+\varepsilon^2/9+O(\varepsilon^3)),\ \ \ 
\hbar\stackrel{\circ}{\omega}_0 \approx 41A^{-1/3} \ {\rm [MeV]}.
\end{equation}
For comparison we will introduce two types of iso-vector dipole 
interaction, one is an ordinary type and the other is a 
doubly-stretched type. The interaction in the latter case is given by 
\begin{equation}\label{IVint}
V_{\lambda=1}^{(T=1)}=-{1 \over 2}\sum\limits_K \chi_{1K}^{(T=1)}
{\left( {Q''_{1K}\tau_z} \right)^{\dag} \left( {Q''_{1K}\tau_z} 
\right)}
\end{equation}
and we will parametrize the force strength as 
$\chi_{1K}^{(T=1)}=-\xi \chi_1^{self}$, where $\chi_1^{self}$ is 
the self-consistent strength of the iso-scalar dipole interaction. 
The typical value of $\xi$ estimated from the symmetry energy 
term in the mass formula under the Fermi gas approximation is about 3 
\cite{BM75}.  

Under the RPA, the excitation energy of each K-component of the GDR 
is obtained analytically as
\begin{eqnarray}
\Omega_{11}&=&\Omega_{GDR}\left({1+{\varepsilon \over {3(1+\xi)}}} 
              \right), \nonumber\\ 
\Omega_{10}&=&\Omega_{GDR}\left({1-{{2\varepsilon} \over {3(1+\xi)}}} 
              \right),
\end{eqnarray}
for the ordinary interaction, while
\begin{equation}
\Omega_{11}=\Omega_{GDR}\left( {1+\varepsilon /3} \right),\ \ \ \ \ 
\Omega_{10}=\Omega_{GDR}\left( {1-2\varepsilon /3} \right),
\end{equation}
for the doubly-stretched interaction. Here $\Omega_{GDR}$ is the 
resonance energy of a spherical nucleus given by
\begin{equation}
\Omega_{GDR}=\sqrt{1+\xi}\omega_0,
\end{equation}
which, in both cases, is compatible with the experimental 
systematics of $\Omega_{GDR} \approx 80A^{-1/3}$ [MeV] if we put 
$\xi=3$. For the doubly-stretched interaction, the total energy 
splitting between the K=0 and K=1 components of the GDR is 
$\varepsilon$ in units of $\Omega_{GDR}$ and is independent of 
$\xi$, which is consistent with the simple classical geometrical 
relation of $(\omega_\bot - \omega_z)/\omega_0=\varepsilon$.
This is in good agreement with the systematics of the 
experimental observation \cite{CBB74,Da58,Ok58}. On the other hand, 
for the ordinary interaction, the splitting is too small by a factor 
of 4 for $\xi=3$. Thus the doubly-stretched interaction model seems 
much more improved than the ordinary one also for the iso-vector 
dipole mode.

%---------------------------------------------------------------------
\subsection{Simple model analysis of GDR of superconductive nuclei}

Effects of the inclusion of the pairing correlation on the properties 
of giant resonances of superconductive nuclei have been studied by 
several authors \cite{PS77,AGG91,CDR91}, and the shifts in excitation 
energies, the changes in energy-weighted sum rule (EWSR), etc., have 
been observed. However as indicated in Refs. \cite{AGG91,CDR91}, some 
of them seem to be spurious due to the violation of the Galilean 
invariance of the system. Here, by use of a simple schematic model, 
we will verify the effects of the pairing correlations on the 
structure of GDR, and will show that such spurious effects can be 
remedied by including the dipole pairing interaction.

The model Hamiltonian is assumed to be 
\begin{equation}\label{HGDR}
H=H_0+V_{\lambda =1}^{(T=1)}+\sum\limits_{\tau}
\left\{ V_{pair}-\lambda \hat{N}+H_{\lambda =1}^{(p)} \right\}_\tau,
\end{equation}
where $H_0$ is the oscillator Hamiltonian of Eq.(\ref{defHO}),
$V_{\lambda=1}^{(T=1)}$ is the iso-vector dipole interaction of 
Eq.(\ref{IVint}), $V_{pair}$ is the monopole pairing potential of 
Eq.(\ref{Vpair}), and $H_{\lambda =1}^{(p)}$ is the dipole pairing 
interaction of Eq.(\ref{Hpair}). 
Here and in the following, the summation index $\tau$ is to be taken 
over the proton and the neutron.
The force strength of the iso-vector dipole interaction is fixed as 
$\xi=3$, while that of the dipole pairing interaction is set to be 
its self-consistent value when it is included.
We perform quasiparticle RPA calculations for a schematic model 
system with N=Z=20.
It must be noticed that the purpose here is not to compare with 
the experimental data of ${}^{40}Ca$ but to investigate the 
fundamental properties of the dipole pairing interaction from the 
purely theoretical point of view.

Table \ref{table1} shows our results on the electric dipole strength 
distributions calculated by assuming the system to be spherical. 
To study the effect of the pairing correlation, we artificially 
change the value of $\Delta$.
For the simplest case of $\Delta=0$ (a), the GDR is located at 24.0 
MeV, and the energy-weighted E1 transition strength of this state 
exhausts its classical sum rule value of
\begin{equation}
S(E1)_{class}={9 \over {4\pi}}{{\hbar^2} \over {2M}}{{NZ} \over A}e^2.
\end{equation}
When the pairing gap is set to be $\Delta=1.0$ MeV (b), reflecting 
the situation that the quasiparticle states are no more degenerate, 
the GDR splits in spite of the assumption that the system is 
spherical. There are mainly three components of 
the GDR ( 21.4, 23.6 and 27.5 MeV). The center of the GDR shifts 
upward about 1 MeV compared to the case of $\Delta=0$, and the EWSR 
is overestimated by about 10\% relative to the classical one. 
In order to remedy this situation, we switch on the dipole 
pairing interaction (c).  In this case, the E1 transition strength 
concentrates again on a single state at about 24.0 MeV, and the EWSR 
recovers to its classical value. 
The fact that the pairing correlation destroys the order in the 
structure of the GDR can further be emphasized by increasing $\Delta$ 
to 2.0 MeV, though the value is not realistic. In this hypothetical 
situation (d), the transition strength splits into mainly 
four components at 20.8, 23.6, 30.0 and 51.3 MeV, and the EWSR 
is overestimated by about 36\%. 
Even in such an extreme situation, if we additionally include the 
dipole pairing interaction (e), the spurious effect of the pairing 
interaction can be removed and the E1 strength is concentrated on a 
single state at 23.9 MeV to recover the EWSR.

Figure \ref{fig1} shows the corresponding results obtained by 
assuming that the model system is deformed to be axially symmetric 
shape of $\beta=0.4$. The continuous strength function, representing 
the transition strength per unit energy, is constructed by using 
the Lorentzian weight function
\begin{equation}\label{Lorent}
\rho (\omega-\omega_\nu)={2 \over \pi}{{\Gamma \omega^2} \over 
{(\omega^2-\omega_\nu^2)^2+\Gamma^2\omega^2}}.
\end{equation}
In the present model calculations, we choose the width to be 
$\Gamma=1.0$ MeV only for the sake of not wiping out the fine 
structure of the resonance. Because of the deformation of the 
system, the GDR splits into K=0 and K=1 components. 
For the case of (a), the GDR shows rather complicated structure. 
The reason for it can be traced to the violation of the Galilean 
invariance for the pairing field and we can eliminate such a 
spurious effect by restoring the broken invariance.
In fact, if we additionally include the dipole pairing interaction, 
the structure of the resonance becomes simpler both for K=0 and K=1 
modes (b). Furthermore, the total energy splitting between the K=0 
and K=1 modes, relative to the average energy of these modes, becomes 
approximately equal to $\beta$ which is consistent to the simple 
classical geometrical relation explained before.

%%%%%%%%%%%%%%%%%%%%%%%%%%%%%%%%%%%%%%%%%%%%%%%%%%%%%%%
\section{Numerical Results}\label{sect5}

In this section, we report some characteristic results of numerical 
calculations in RPA of the GDR's of ${}^{148,154}Sm$, the first 
excited $2^+$ states of Sn isotopes and the first excited $3^-$ 
states of Mo isotopes. It should be noted here that 
our present calculations contain essentially 
only one free parameter in the following sense; 
the single particle bases are 
constructed from the Nilsson + BCS model with standard parameters; 
the strengths of the velocity dependent effective interactions such as 
the multipole pairing interactions are 
fixed to be their self-consistent values when they are included; 
only the strengths of the multipole-multipole interactions 
are adjusted under the condition that a common value 
of $\xi$, $\chi_2$ and $\chi_3$ 
is to be adopted for all the isotopes of Sm, Sn and Mo, respectively.

Since the positions of band-head states are very sensitive to the 
choice of the single particle energies as are generally observed in the 
RPA calculation of vibrational states in deformed nuclei, one can improve 
the fit to experiments by adjusting the single-particle energies, and 
further improvement can be obtained by slightly varying the strengths 
of the effective interactions around the vicinity of the predicted 
self-consistent values for each isotope. The fit to experiment 
we obtained is insufficient and an improved fit 
could have been obtained if these parameters were treated as adjustable 
as well. We have not done so in the present work, because 
the purpose of the present numerical investigation is to see first 
to what extent our theory works without much playing 
around with parameters and to provide understandings rather than 
precision tools for the fitting of experimental data.

%---------------------------------------------------------------------
\subsection{E1 strength distributions of ${}^{148,154}$Sm}

We here present the results of realistic calculations on the E1 
strength distributions of ${}^{148,154}$Sm in the quasiparticle RPA. 
The model Hamiltonian is assumed to be 
\begin{equation}\label{Nil}
H=H_0+V_{Nil}+V_{\lambda =1}^{(T=1)}+\sum\limits_{\tau}
\left\{ V_{pair}-\lambda \hat{N}+H_{\lambda =1}^{(p)} \right\}_\tau,
\end{equation}
\begin{equation}
V_{Nil}=\sum_{i=1}^A \left[-\kappa \hbar\stackrel{\circ}{\omega}_0 
\left\{ 2(\vec l \cdot \vec s)
+\mu(\vec l{}^2 -<\vec l{}^2>)\right\} \right]_i,
\end{equation}
which is essentially same as Eq.(\ref{HGDR}) except that the 
Nilsson potential $V_{Nil}$ is additionally included.

We will study the effects of two kinds of velocity-dependent 
interactions to restore the Galilean invariance of the system; 
the one arising from $V_{pair}$ and the other from the 
velocity dependent part of $V_{Nil}$. The former, the dipole pairing 
interaction $H_{\lambda =1}^{(p)}$, is given by Eq.(\ref{Hpair}), 
while the latter, $H_{\lambda =1}^{(Nil)}$, is given by
\begin{equation}
H_{\lambda =1}^{(Nil)}={1 \over 2}\sum\limits_\mu \kappa_1 
\eta_{1\mu}^{\dag} \eta_{1\mu}
\end{equation}
with
\begin{equation}
\eta_{1\mu}=[Q''_{1\mu},V_{Nil}] ,
\end{equation}
\begin{equation}
\kappa_1=-1/< [Q''_{1\mu},[V_{Nil},Q''_{1\mu}{}^{\dag}]]>_0 .
\end{equation}
Although it is known that there exists a term associated with the 
coordinate distortion of the spin-orbit potential given by Eq.(6-70) 
of Ref.\cite{BM75}, we will not consider its effects in order to 
concentrate our present analysis on the effects of the 
self-consistent 
velocity dependent interactions. We will also neglect iso-vector 
corrections coming from the Nilsson potential to the iso-vector 
dipole-dipole interaction because of the same reason.

To fix the parameter $\xi$ for the strength of the 
iso-vector dipole-dipole interaction in this mass region,
we first calculate the case of ${}^{148}$Sm by assuming its shape 
to be spherical. The resultant value of $\xi=2.7$ is adopted also 
for ${}^{154}$Sm whose quadrupole deformation parameter is assumed 
and fixed as $\beta=0.35$ in the following calculations. All other 
strengths of the interactions are fixed just as the self-consistent 
values which restore the Galilean invariance of the system. 
It should be noticed here that the doubly-stretched interactions 
are used for the deformed nucleus of ${}^{154}$Sm.
For the model space, we retain all the Nilsson single particle 
states with $2\leq N_{osc}\leq 7$ for protons and 
$3\leq N_{osc}\leq 8$ for neutrons. 
The Nilsson parameters are taken from  Ref.\cite{NTS69}. 
By using experimental binding energies of Ref.\cite{WA85},
the pairing gap parameters are determined from the even-odd mass 
differences as $\Delta_n=1.01$ MeV, $\Delta_p=1.36$ MeV for 
${}^{148}Sm$ and $\Delta_n=1.07$ MeV, $\Delta_p=0.86$ MeV for 
${}^{154}Sm$. We use the Lorentzian distribution of Eq.(\ref{Lorent}) 
to reproduce the resonance width.  We choose $\Gamma$ (in units of 
MeV) as 5.10 for ${}^{148}Sm$, and 3.25, 5.25 for the K=0, 1 modes of 
${}^{154}Sm$, respectively.

Figures \ref{fig2}a and \ref{fig2}b show the E1 strength functions 
for ${}^{148}Sm$ and ${}^{154}Sm$, respectively, 
taking the model Hamiltonian of Eq.(\ref{Nil}) but 
switching off the dipole-pairing interaction.
Here we see that even if we use the doubly-stretched interaction, 
the calculated splitting between K=0 and K=1 resonances of 
${}^{154}Sm$ is too small. Furthermore, as can be seen from the  
column a of Table \ref{table2}, the EWSR values exceed the classical 
values for both nuclei.  
As stated repeatedly, these difficulties stem from the mixture of 
spurious states arising from the broken Galilean invariance of 
$V_{pair}$ and $V_{Nil}$. In the following we show the results 
obtained by restoring the broken symmetry in two step.

First, we study the effect of $H_{\lambda=1}^{(p)}$. The column b 
of Table \ref{table2} shows that the sum rule values approach 
the classical values. However, from Fig.\ref{fig3}, we see that 
the centers of the resonances shift to lower excitation energies 
and the structure in the lower peak region reveal unnatural shape. 
Finally we take into account $H_{\lambda=1}^{(Nil)}$ so that the 
Galilean invariance of the Hamiltonian is restored. As can be 
seen from Fig.\ref{fig4} and column c of Table \ref{table2}, the 
unnatural resonance structure disappears and the EWSR values keep 
close to the classical limit.
We note that the final results of Figs.\ref{fig4}a and \ref{fig4}b 
agree quite well with the Lorentzian distribution of Figs.\ref{fig4}c 
and \ref{fig4}d which fit the experimental data of photoneutron cross 
section \cite{CBB74}.

%---------------------------------------------------------------------
\subsection{The first excited $2^+$ states in Sn isotopes}

We now consider the effects of the self-consistent quadrupole pairing 
interaction on the excitation energies $E(2_1^+)$ and the E2 
transition probabilities B(E2) of Sn isotopes within the 
quasiparticle RPA. The model Hamiltonian is assumed to be 
\begin{equation}\label{HSn}
H=H_0+V_{Nil}+V_{\lambda =2}+\sum\limits_{\tau}
\left\{ V_{pair}-\lambda \hat{N}+H_{\lambda =2}^{(p)} \right\}_\tau,
\end{equation}
where $V_{\lambda =2}$ and $H_{\lambda =2}^{(p)}$ are the 
quadrupole-quadrupole interaction and the quadrupole pairing 
interaction, respectively.

The single particle model space is spanned by all the Nilsson 
states with $2\leq N_{osc}\leq 6$ for protons and 
$2\leq N_{osc}\leq 7$ for neutrons. The Nilsson parameters are 
taken from  Ref.\cite{NTS69}, and the deformation is set to be zero. 
To investigate the effect of the quadrupole pairing force on the 
$B(E2;0_g \rightarrow 2_1^+)$ value, the strength of the quadrupole 
pairing force, $G_2$, is fixed to be its self-consistent value 
when it 
is included, while that of the quadrupole interaction, $\chi_2$, is 
used as an adjustable parameter to reproduce experimental 
$E(2_1^+)$ of ${}^{116,118,120}Sn$ \cite{LS78}.

First, we adopt the pairing gaps determined from the even-odd mass 
differences. The adopted value of $\chi_2$ for the best fit 
obtained without (with) the inclusion of the quadrupole-pairing 
interaction, $H_{\lambda=2}^{(p)}$, is 0.92 (0.90) in units of 
$\chi_2^{self}$. The calculated $E(2_1^+)$ and B(E2) are plotted in 
Figs.\ref{fig5}a and \ref{fig5}b, respectively, and corresponding 
numerical values are given in column a of Table \ref{table3}. 
These results show that $H_{\lambda=2}^{(p)}$ is necessary and 
important in reproducing experimental $E(2_1^+)$ and B(E2) values 
simultaneously. 

Second, we have also performed numerical calculations by fixing the 
proton energy gap to be zero. This is because the proton shell is 
closed in Sn. In this case, the adopted value of $\chi_2$ for the 
best fit obtained without (with) the inclusion of 
$H_{\lambda=2}^{(p)}$ is 0.940 (0.965) in units of $\chi_2^{self}$.
The calculated values of $E(2_1^+)$ and B(E2) are plotted in 
Figs.\ref{fig6}a and \ref{fig6}b, respectively, and corresponding 
numerical values are given in column b of Table \ref{table3}, 
where we see better agreements with experimental data compared to the 
case with $\Delta_p \ne 0$.

%---------------------------------------------------------------------
\subsection{The first Excited $3^-$ States in Mo Isotopes}

Here we will study the effects of the self-consistent octupole 
pairing interaction, $H_{\lambda=3}^{(p)}$, in the quasiparticle RPA 
calculations of the excitation energies $E(3_1^-)$ and the E3 
transition probabilities of the first excited $3^-$ states in Mo 
isotopes. The model Hamiltonian is assumed to be 
\begin{equation}\label{HMo}
H=H_0+V_{Nil}+V_{\lambda =3}+\sum\limits_{\tau}
\left\{ V_{pair}-\lambda \hat{N}+H_{\lambda =3}^{(p)} \right\}_\tau,
\end{equation}
where $V_{\lambda =3}$ and $H_{\lambda =3}^{(p)}$ are the 
octupole-octupole interaction and the octupole pairing 
interaction, respectively.

The single particle model space is spanned by all the Nilsson 
states with $0\leq N_{osc}\leq 7$ for protons and 
$0\leq N_{osc}\leq 9$ for neutrons. Energy gaps are 
determined from the experimental even-odd mass differences. 
To investigate the effect of the octupole pairing force on the 
$B(E3;0_g \rightarrow 3_1^-)$ value, the strength of the octupole 
pairing force, $G_3$, is fixed to be its self-consistent value 
when it 
is included, while that of the octupole interaction, $\chi_3$, is 
used as an adjustable parameter to reproduce the 
experimental data of $E(3_1^-)$ in ${}^{94,96,98,100}Mo$ \cite{BBB72}.
The adopted value of $\chi_3$ for the best fit 
obtained without (with) the inclusion of the octupole-pairing 
interaction, $H_{\lambda=3}^{(p)}$, is 0.95 (0.97) in units of 
$\chi_3^{self}$. 

The calculated $E(3_1^-)$ and B(E3) from the ground state to the 
$3_1^-$ state are plotted in 
Figs.\ref{fig7}a and \ref{fig7}b, respectively, and corresponding 
numerical values are given in Table \ref{table4}.
Here we see that the calculated B(E3) values are reduced and improved 
systematically by the effects of $H_{\lambda=3}^{(p)}$, although the 
fit to experimental data is insufficient.

%%%%%%%%%%%%%%%%%%%%%%%%%%%%%%%%%%%%%%%%%%%%%%%%%%%%%%%
\section{Summary}\label{sect6}

We have extensively applied the prescription to derive 
self-consistent effective interactions, needed for the unified 
description of the collective motions of atomic nuclei, 
especially to the system with velocity dependent mean potentials.
As the guiding principles for this purpose we have imposed the 
conditions of nuclear self-consistency \cite{KMY75,SK89} and 
local Galilean invariance \cite{BM75,Be67,PS77,PGS77,SK90} of the 
system, implemented with the simple and transparent field 
coupling method developed by Bohr and Mottelson \cite{BM75}.

The nuclear self-consistency requires that the shape of the mean 
potential and that of the density are the same even when the system 
undergoes collective motions, while the local Galilean invariance 
requires that the nucleonic velocities entering into the 
velocity-dependent single-particle potential should be expressed 
relative to the local collective flow. In the field coupling method, 
the coupling between the particle motion 
and the collective field is identified, within the Hartree 
approximation, as the averaged one-body field of the effective 
interaction which we look for, and the coupling strength of it can 
be fixed by the above conditions. 

For the multipole collective shape oscillation modes in the harmonic 
oscillator potential with the monopole pairing correlation, we have 
derived the multipole-multipole interactions for particle-hole 
channel and the multipole-pairing interactions for particle-particle 
channel from the unified physical picture. 
In the case of deformed nuclei, it is shown that these interactions 
must be expressed in terms of the doubly-stretched coordinates so as 
to guarantee the conditions of nuclear self-consistency and local 
Galilean invariance of the system. 

The origin of the doubly-stretched multipole-multipole interactions 
have already been clarified and they have found many successful 
applications 
\cite{KMY75,SK89,KT72,KT76,TWK79,SK88,SK91,MM87,Ai90,MSM90}, 
while the origin of the doubly-stretched multipole pairing 
interactions are clarified in this paper on the same footing. 
We have applied the doubly-stretched multipole pairing 
interactions to the analyses of some collective 
states in Sm, Sn and Mo isotopes by means of RPA, 
and for the dipole mode we have also tested the velocity 
dependent effective interaction arising from the Nilsson potential. 
We have seen the effects of such velocity dependent effective 
interactions in the recovery of the 
classical E1 sum rule for GDR's of ${}^{148,154}Sm$, 
in the systematic reduction of the E2 transition probabilities 
for the first excited $2^+$ states of Sn isotopes, 
and also in the systematic reduction of E3 transition probabilities 
for the first excited $3^-$ states of Mo isotopes.
It should be noted here that recently the doubly-stretched 
quadrupole pairing interaction was successfully applied also to 
the microscopic analysis of identical bands in superdeformed nuclei, 
and it was shown that the doubly-stretched 
quadrupole pairing interaction has several advantages compared 
to the non-stretched and stretched ones \cite{SW94}.

In summary, it is clarified that the self-consistent velocity 
dependent effective interactions play crucial roles 
to recover the local Galilean invariance and eliminate 
various unphysical effects arising from the spurious velocity 
dependence of the mean potential. 
For rotating nuclei, we can apply similar prescription in order 
to find the proper effective interactions which faithfully take 
into account the effects of the collective rotation. 
Results of it will be reported in a separate paper.

%---------------------------------------------------------------------
\acknowledgments

We are very much grateful to Dr. S.-I. Kinouchi for valuable 
discussions through the course of the present work. 
One of the authors (T. Kubo) 
would like to express his sincere thanks to Dr. T. Marumori, 
Dr. K. Matsuyanagi and Dr. F. Sakata for helpful comments and continuous 
encouragement. He also acknowledges Dr. O. Morimatsu and the 
late Dr. Y. Miyama for their help in computations.
One of the authors (H.S.) would like to express his gratitude to 
Dr. R. Wyss for the useful comments on the applications of the 
doubly-stretched quadrupole pairing interaction. 

%---------------------------------------------------------------------

%---------------------------------------------------------------------
\begin{figure}
\caption{\label{fig1} E1 transition strength distribution for a 
schematic model system of N=Z=20 with axially symmetric deformation 
of $\beta=0.4$ calculated in RPA. The continuous strength function, 
representing the strength per unit energy, is given in units of 
MeV${}^{-1}$ relative to the classical sum rule value (CSR). In the 
model Hamiltonian of Eq.(\protect\ref{HGDR}), the pairing gap is 
fixed as $\Delta=2.0$ MeV both for protons and for neutrons. K=0 and 
K=1 modes are shown by the solid and the dashed curves, respectively. 
(a) and (b) correspond to the results obtained with and without the 
inclusion of the dipole-pairing interaction, respectively.}
\end{figure}

\begin{figure}
\caption{\label{fig2} E1 transition strength functions of 
(a) ${}^{148}$Sm and (b) ${}^{154}$Sm are given in units of 
MeV${}^{-1}$ relative to the classical sum rule value (CSR). 
The model Hamiltonian is same 
as that of Eq.(\protect\ref{Nil}) except that the dipole-pairing 
interaction is not included here. In (b), solid and dashed curves 
correspond to the K=0 and K=1 modes, respectively.}
\end{figure}

\begin{figure}
\caption{\label{fig3} Same as Fig.\protect\ref{fig2}, except that the 
dipole-pairing interaction is included.}
\end{figure}

\begin{figure}
\caption{\label{fig4} Same as Fig.\protect\ref{fig3} for (a) and (b), 
except that the additional interaction $H_{\lambda=1}^{(Nil)}$ is 
included. For comparison, Lorentzian distributions which fit the 
experimental data of the photoneutron cross section \protect\cite{CBB74} 
are given in (c) and (d) for ${}^{148}$Sm and ${}^{154}$Sm, 
respectively. The experimental Lorentz line parameters 
for the mathematical expression 
$\sigma_L(E)=\sum_{i}\sigma_i {{(E\Gamma_i)^2} \over 
{(E^2-E_i^2)^2+(E\Gamma_i)^2}}$ 
are taken from Ref.\protect\cite{CBB74} as $E_1=14.8 \pm 0.1$, 
$\Gamma_1=5.1 \pm 0.2$, $\sigma_1=339 \pm 12$ for the best single 
line fit of ${}^{148}Sm$ and $E_1=12.35 \pm 0.10$, 
$\Gamma_1=3.35 \pm 0.15$, $\sigma_1=192 \pm 10$, $E_2=16.1 \pm 0.1$, 
$\Gamma_2=5.25 \pm 0.20$, $\sigma_2=204 \pm 10$ for the best two line 
fit of ${}^{154}Sm$, respectively. Here $E$'s and $\Gamma$'s are 
given in units of MeV while $\sigma$'s are given in units of mb.}
\end{figure}

\begin{figure}
\caption{\label{fig5} (a) Excitation energies of the first $2^+$ 
states and (b) E2 transition probabilities in Sn isotopes. Results of 
calculations with and without the inclusion of the quadrupole pairing 
interaction are shown by dot-dashed and dashed lines, respectively. 
Solid lines correspond to the experimental data. In the 
calculations, pairing gaps are determined from the odd-even mass 
differences both for protons and for neutrons.}
\end{figure}

\begin{figure}
\caption{\label{fig6} Same as Fig.\protect\ref{fig5}, except that the 
pairing gaps for protons are set to be $\Delta_p=0$ in the 
calculations.}
\end{figure}

\begin{figure}
\caption{\label{fig7} (a) Excitation energies of the first $3^-$ 
states and (b) E3 transition probabilities in Mo isotopes. Results of 
calculations with and without the inclusion of the octupole pairing 
interaction are shown by dot-dashed and dashed lines, respectively.
Solid lines correspond to the experimental data.}
\end{figure}

%---------------------------------------------------------------------
%\widetext
%\mediumtext
\narrowtext
\begin{table}
\caption{\label{table1} E1 transition strength distribution for a 
               schematic spherical system with N=Z=20. Energies and 
               fractions of the EWSR are shown for some dominant states 
               calculated in RPA. The dipole-pairing interaction is 
               included for c and e, but not for a, b, and d.}
\begin{tabular}{cddd}
  & $\Delta_p=\Delta_n$ (MeV) & E (MeV) & EWSR (\%) \\
\tableline
a & 0.0 & 24.0 & 100 \\
\tableline
b & 1.0 & 21.4 &   8 \\
  &     & 23.6 &  52 \\
  &     & 27.5 &  41 \\
\tableline
c & 1.0 & 24.0 & 100 \\
\tableline
d & 2.0 & 20.8 &  16 \\
  &     & 23.6 &  24 \\
  &     & 30.0 &  78 \\
  &     & 51.3 &  13 \\
\tableline
e & 2.0 & 23.9 &  99 \\
\end{tabular}
\end{table}

%\widetext
%\mediumtext
\narrowtext
\begin{table}
\caption{\label{table2} Calculated E1 EWSR values of 
         ${}^{148,154}Sm$. Strengths integrated over the 
         excitation energies from 5.5 MeV to 30.0 MeV are given 
         in units of \% relative to the classical sum rule value. 
         The model Hamiltonian is given by Eq.(\protect\ref{Nil}).
         The columns a and b correspond to the results obtained 
         without and with the inclusion of the dipole-pairing 
         interaction $H_{\lambda=1}^{(p)}$, respectively, while in c 
         the interaction $H_{\lambda=1}^{(Nil)}$ is also included 
         in addition to $H_{\lambda=1}^{(p)}$.}
\begin{tabular}{lddd}
Nucleus  & a & b & c \\
\tableline
${}^{148}Sm$ & 116.8 & 85.9 & 86.2 \\
${}^{154}Sm$ (K=0) & 41.1 & 31.3 & 30.2 \\
${}^{154}Sm$ (K=1) & 64.7 & 56.0 & 56.9 \\
\end{tabular}
\end{table}

%\widetext
\mediumtext
%\narrowtext
\begin{table}
\caption{\label{table3} Energies of the first excited $2^+$ states 
         and E2 transition probabilities in Sn isotopes 
         are given in units of MeV and $B(E2)_{sp}$, respectively.
         Results of calculations obtained without and with the 
         inclusion of the quadrupole-pairing interaction 
         $H_{\lambda=2}^{(p)}$ are compared with experimental data. 
         The pairing gaps adopted in the calculations are taken from 
         the experimental even-odd mass differences for a, while 
         those for protons are set to be zero for b.}
\begin{tabular}{ccccccccccd}
 &\multicolumn{4}{c}{a}&\multicolumn{4}{c}{b}
&\multicolumn{2}{c}{}\\
\cline{2-5}\cline{6-9}
 &\multicolumn{2}{c}{without $H_{\lambda=2}^{(p)}$}
 &\multicolumn{2}{c}{with $H_{\lambda=2}^{(p)}$}
 &\multicolumn{2}{c}{without $H_{\lambda=2}^{(p)}$}
 &\multicolumn{2}{c}{with $H_{\lambda=2}^{(p)}$}
 &\multicolumn{2}{c}{Exp.}\\
\cline{2-3}\cline{4-5}\cline{6-7}\cline{8-9}\cline{10-11}
N&$E(2^+)$&B(E2)&$E(2^+)$&B(E2)&$E(2^+)$&B(E2)
&$E(2^+)$&B(E2)&$E(2^+)$&B(E2)\\
\tableline
62&1.72&18.7&1.57&13.3&1.66&15.7&1.57&10.7&1.26&16.2\\
64&1.72&19.6&1.57&14.1&1.68&15.9&1.58&10.8&1.30&15.3\\
66&1.41&19.8&1.19&17.1&1.37&15.7&1.34&10.6&1.29&12.9\\
68&1.23&23.9&1.03&20.4&1.20&19.0&1.21&12.2&1.23&12.7\\
70&1.20&24.8&1.00&21.2&1.16&20.0&1.18&12.7&1.17&11.6\\
72&1.22&22.7&1.13&15.0&1.17&18.5&1.18&11.9&1.14&10.9\\
74&1.35&18.0&1.27&12.9&1.31&14.9&1.27&10.0&1.13& 9.3\\
\end{tabular}
\end{table}

%\widetext
%\mediumtext
\narrowtext
\begin{table}
\caption{\label{table4} Energies of the first excited $3^-$ states 
         and E3 transition probabilities in Mo isotopes 
         are given in units of MeV and $B(E3)_{sp}$, respectively.
         Results of 
         calculations obtained without and with the inclusion of the 
         octupole-pairing interaction $H_{\lambda=3}^{(p)}$ are 
         compared with experimental data. The pairing 
         gaps adopted in the calculations are taken from the 
         experimental even-odd mass differences.}
\begin{tabular}{ccccccd}
 &\multicolumn{2}{c}{without $H_{\lambda=3}^{(p)}$}
 &\multicolumn{2}{c}{with $H_{\lambda=3}^{(p)}$}
 &\multicolumn{2}{c}{Exp.}\\
\cline{2-3}\cline{4-5}\cline{6-7}
N&$E(3^-)$&B(E3)&$E(3^-)$&B(E3)&$E(3^-)$&B(E3)\\
\tableline
52&2.74&51.2&2.66&44.3&2.53&17\\
54&2.36&53.4&2.32&44.6&2.23&24\\
56&1.93&58.6&1.97&47.3&2.02&33\\
58&1.91&60.2&1.92&47.8&1.91&32\\
\end{tabular}
\end{table}


\begin{references}
\bibitem[\dag]{byline1} Current address: 1-1-34 Minato, Izumisano-shi, 
                        Osaka 598, Japan.
\bibitem[*]{byline2}  Deceased 14 April 1990.
\bibitem{BM75}  A. Bohr and B. R. Mottelson, {\it Nuclear Structure}, 
         (Benjamin, New York, 1975) Vol. II, Chap.6.
\bibitem{Mo67}  B. R. Mottelson, {\it Nikko Summer School Lectures}, 
         (NORDITA, Copenhagen, 1967) pub.288. 
\bibitem{Be67}  S. T. Belyaev, Sov. J. Nucl. Phys. {\bf 4}, 671 (1967).
\bibitem{Ha74}  I. Hamamoto, Nucl. Phys. {\bf A232}, 445 (1974).
\bibitem{PS77}  N. I. Pyatov and D. I. Salamov, Nukleonika, 
         {\bf 22}, 127 (1977).
\bibitem{PGS77} N. I. Pyatov, S. I. Gabrakov and D. I. Salamov, 
         Sov. J. Nucl. Phys. {\bf 26}, 139 (1977).
\bibitem{SK90}  H. Sakamoto and T. Kishimoto, Phys. Lett. 
         {\bf 245B}, 321 (1990).
\bibitem{KSK88} T. Kishimoto, H. Sakamoto, S.-I. Kinouchi, T. Kubo 
         and T. Kammuri, in {\it Proceedings of the Texas A\&M 
         Symposium on Hot Nuclei}, edited by S. Shlomo, 
         R. P. Schmitt and J. B. Natowitz (World Scientific, 
         Singapore, 1988) p.89.
\bibitem{KSK89} S.-I. Kinouchi, H. Sakamoto, T. Kubo and T. Kishimoto,
         in {\it Nuclear Collective Motion and Nuclear 
         Reaction Dynamics}, edited by K.-I. Kubo, 
         M. Ichimura, M. Ishihara and S. Yamaji 
         (World Scientific, Singapore, 1989) p.111.
\bibitem{KMY75} T. Kishimoto, J. M. Moss, D. H. Youngblood, J. D.          
            Bronson, C. M. Rozsa, D. R. Brown and A. D. Bacher, 
         Phys. Rev. Lett. {\bf 35}, 552 (1975).
\bibitem{SK89}  H. Sakamoto and T. Kishimoto, Nucl. Phys.
         {\bf A501}, 205 (1989); {\bf A501}, 242 (1989).
\bibitem{Ni55}  S. G. Nilsson, Mat. Fys. Medd. Dan. Vid. Selsk. 
         {\bf 29} (1955) No.16.
\bibitem{NTS69} S. G. Nilsson, C. F. Tsang, A. Sobiczewski, 
         Z. Szyma\'{n}ski, S. Wycech, C. Gustafson, 
         I.-L. Lamm, P.M\"{o}ller and B. Nilsson, 
         Nucl. Phys. {\bf A131}, 1 (1969).
\bibitem{KTK83} T. Kishimoto, T. Tamura and T. Kammuri, 
         Prog. Theor. Phys. Suppl. {\bf 74\&75}, 170 (1983).
\bibitem{Ma83}  E. R. Marshalek, 
         Phys. Rev. Lett. {\bf 51}, 1534 (1983); 
                Phys. Rev. C {\bf 29}, 640 (1984); 
                Phys. Lett. {\bf 244B}, 1 (1990).
\bibitem{KT72}  T. Kishimoto and T. Tamura, 
         Nucl. Phys. {\bf A192}, 246 (1972).
\bibitem{KT76}  T. Kishimoto and T. Tamura, 
         Nucl. Phys. {\bf A270}, 317 (1976).
\bibitem{TWK79} T. Tamura, K. J. Weeks and T. Kishimoto, 
         Phys. Rev. C {\bf 20}, 307 (1979); 
         Nucl. Phys. {\bf A347}, 359 (1980).
\bibitem{SK88}  H. Sakamoto and T. Kishimoto, 
         Nucl. Phys. {\bf A486}, 1 (1988).
\bibitem{SK91}  H. Sakamoto and T. Kishimoto, 
         Nucl. Phys. {\bf A528}, 73 (1991).
\bibitem{MM87}  M. Matsuo and K. Matsuyanagi, 
         Prog. Theor. Phys. {\bf 78}, 591 (1987).
\bibitem{Ai90}  H. Aiba, Prog. Theor. Phys. {\bf 84}, 908 (1990).
\bibitem{MSM90} S. Mizutori, Y. R. Shimizu and K. Matsuyanagi, 
         Prog. Theor. Phys. {\bf 83}, 666 (1990); 
                              {\bf 85}, 559 (1991).
\bibitem{AGG91} J. M. Arias, M. Gallardo and J. G\'{o}mez-Camacho, 
         Nucl. Phys. {\bf A528}, 144 (1991).
\bibitem{CDR91} O. Civitarese, A. G. Dumrauf, M. Reboiro, P. Ring and 
         M. M. Sharma, Phys. Rev. C {\bf 43}, 2622 (1991).
\bibitem{CKC84} S. Cwiok, J. Kvasil and B. Choriev,
         J. Phys. {\bf G10}, 903 (1984).
\bibitem{NF88}  R. Nojarov and A. Faessler, 
         Nucl. Phys. {\bf A484}, 1 (1988).
\bibitem{CL88}  O. Civitarese and M. C. Licciardo, 
         Phys. Rev. C {\bf 38}, 967 (1988); 
         {\bf 39}, 1550 (1989).
\bibitem{CFL92} O. Civitarese, A. Faessler and M. C. Licciardo, 
         Nucl. Phys. {\bf A542}, 221 (1992).
\bibitem{NFD95} R. Nojarov, A. Faessler and M. Dingfelder, 
         Phys. Rev. C {\bf 51}, 2449 (1995).
\bibitem{Wi32}  E. Wigner, Phys. Rev. {\bf 40}, 749 (1932).
\bibitem{RS80}  P. Ring and P. Schuck, {\it The Nuclear Many-Body 
                Problem}, (Springer, New York, 1980).
\bibitem{BS80}  R. Bengtsson and P. Schuck, 
         Phys. Lett, {\bf 89B}, 321 (1980).
\bibitem{DK87}  M. Di Toro and V. M. Kolomietz, 
         Z. Phys. A {\bf 328}, 285 (1987).
\bibitem{CBB74} P. Carlos, H. Beil, R. Bergere, A. Lepretre, 
         A. de Miniac and A. Veyssiere, 
         Nucl. Phys. {\bf A225}, 171 (1974).
\bibitem{Da58}  M. Danos, Nucl. Phys. {\bf 5}, 23 (1958).
\bibitem{Ok58}  K. Okamoto, Phys. Rev. {\bf 110}, 143 (1958).
\bibitem{WA85}  A. H. Wapstra and G. Audi, 
         Nucl. Phys. {\bf A432}, 1 (1985).
\bibitem{LS78}  C. M. Lederer and V. S. Shirley, 
         {\it Table of Isotopes} (7th edition), 
         (John Wiley \& Sons, New York, 1978).
\bibitem{BBB72} J. Barrette, M. Barrette, A. Boutard, R. Haroutunian,
         G. Lamoureux and S. Monaro, 
         Phys. Rev. C {\bf 6}, 1339 (1972).
\bibitem{SW94}  W. Satu{\l}a and R. Wyss, 
         Phys. Rev. C {\bf 50}, 2888 (1994).
\end{references}
\end{document}